\documentstyle[12pt,aaspp4]{article}

\def\NH3{NH$_3$}
\def\C18O{C$^{18}$O}
\def\km/s{km\,s$^{-1}$}
\def\N{$N$}
\def\msarcsec{m\,s$^{-1}$\,arcsec$^{-1}$}
\def\msAU{m\,s$^{-1}$\,AU$^{-1}$}
\def\Jb{Jy\,beam$^{-1}$}
\def\Jbk{Jy\,beam$^{-1}$\km/s}

\begin{document}

\title{33.8 GHz CCS Survey of Molecular Cores in Dark Clouds}
\author{Shih-Ping Lai \& Richard M. Crutcher}
\affil{Department of Astronomy, University of Illinois, Urbana, IL 61801;\\
 slai@astro.uiuc.edu, crutcher@astro.uiuc.edu}
 



\begin{abstract}
 
We have conducted a survey of the CCS $J_N = 3_2-2_1$ line
toward 11 dark clouds and star-forming regions
at 30\arcsec~spatial resolution and 0.054 \km/s velocity resolution.
CCS was only detected in 
quiescent clouds, not in active star-forming regions.  
The CCS distribution shows remarkable clumpy structure,
and 25 clumps are identified in 7 clouds.
Seven clumps with extremely narrow nonthermal linewidths $<$ 0.1 \km/s
are among the most quiescent clumps ever found.
The CCS clumps tend to exist 
around the higher density regions traced by \NH3 emission or 
submillimeter continuum sources, and
the distribution is not spherically symmetric.
Variation of the CCS abundance was suggested as
an indicator of the evolutionary status of star formation.
However, we can only find a weak correlation between \N(CCS) 
and $n_{H_2,vir}$.
The velocity distributions of CCS clouds reveal that
a systematic velocity pattern generally exists.
The most striking feature in our data is a ring structure in the 
position-velocity diagram of L1544 with an well-resolved
inner hole of 0.04 pc\,$\times$\,0.13 
\km/s and an outer boundary of 0.16 pc\,$\times$\,0.55 \km/s.  
This position-velocity structure clearly indicates an edge-on disk or ring 
geometry, and it can be interpreted as a collapsing disk with 
an infall velocity $\gtrsim$ 0.1 \km/s and a 
rotational velocity less than our velocity resolution.
Nonthermal linewidth distribution is generally coherent in CCS clouds,
which could be evidence for the termination of Larson's Law 
at small scales, $\sim$ 0.1 pc.

\end{abstract}

\clearpage

%
%

\section{Introduction}

The CCS molecule was first discovered toward TMC-1 by Suzuki et al.\ (1984)
and was identified in the laboratory by Saito et al.\ (1987).  It is a very
interesting molecule because it is abundant in starless clouds
(Suzuki et al.\ 1992, hereafter SYO) with an appropriate critical density 
($10^{4}-10^{5}$ cm$^{-3}$) for pre-protostellar cores
(Wolkovitch et al.\ 1997, hereafter WLG).
CCS is also an excellent probe of kinematic structure
of dark clouds because of its lack of hyperfine splittings and
its moderate optical depth.  Moreover, 
the CCS structure is Hund's case b and as such has a large Zeeman splitting
factor of 0.70~Hz\,$\mu$G$^{-1}$ at 33 GHz (Shinnaga \& Yamamoto, 1999), 
which provides an opportunity to explore the magnetic field structure
in the early stage of star formation.  

Several single-dish surveys of CCS have been done in the past 10 years.
The survey done by Fuente et al.\ (1990) showed that 
N(CCS) increases with N(\C18O) and N(HC$_5$N).  
However, this correlation is not so obvious for later observations (\S 4.2).
SYO performed the largest single-dish survey 
toward 49 dark cores and concluded that
CCS is more abundant in cold quiescent dark clouds while less
abundant in star-forming regions.
They also constructed a pseudo-time-dependent chemical evolution model,
suggesting that CCS is more abundant in the early stage
of chemical evolution whereas \NH3 is more abundant in the later stage.
[CCS]/[\NH3] was proposed as an indicator of cloud evolution
and star formation.
Based on this suggestion, Scappini \& Codella (1996)
observed \NH3 and CCS toward a sequence of objects in different evolutionary
stages -- dark clouds, inactive Bok globules, active Bok globules, and
massive protostars.  However, because of the large scatter, 
their data did not determine a trend.   
The most recent survey done by Benson et al.\ (1998)
only showed a slight tendency for CCS to have a higher column density in
starless cores, but their result was not statistically significant.

CCS maps of dark clouds have been made toward B335 and L1498.
Velusamy et al.\ (1995) combined VLA and DSN 70\,m single-dish maps of 
the 22 GHz CCS $J_N=2_1-1_0$ transition toward B335 
and showed that the distribution of CCS emission 
is non-spherical and clumpy with depletion in the central 10\arcsec.
They concluded that CCS is characteristic of early astrochemical time
(t$\le 10^5$ yr) and more prominent in the region
just outside the infall envelope.  WLG showed that
the CCS emission of L1498 has two peaks separated 
by 2\arcmin~ in the SE-NW direction. However, the NH$_3$ map (\cite{bm89})
only shows a single component between two CCS components.
Kuiper et al.\ (1996) interpreted this as the CCS
originating in a shell outside of the ammonia core,
and concluded that L1498 may be about to
collapse rapidly to form a protostar.

In this paper, we present a survey with largest number of
CCS maps and spectra.  These data with large field of view ($\sim$
6\arcmin) and adequate velocity and spatial resolutions provide 
a better foundation to examine the existent arguments.
In \S2 we describe our observation and data reduction procedure.  
In \S3 we present the analysis for obtaining physical parameters and
discuss the results in \S4.  We summarize the conclusions in \S5.

\section{Observations}

We used 1-cm receivers (Carlstrom 1996) 
on the Berkeley-Illinois-Maryland Association (BIMA, Welch et al 1996) array
to observe the CCS $J_N=3_2-2_1$ line toward 11 dark clouds or star-forming
regions.  The sources we observed are listed in Table 1.  
The frequency of this CCS transition is 33.75138 GHz (WLG).

Most of the observations were carried out between 1997 Aug 27 and Sep 3
with one additional track for extended B1 emission on 1998 Sep 7.
During the observations in 1997, BIMA was in the D configuration 
with 9 antennas.
The range of the baselines of D configuration was 0.6--8.0 kilowavelength, 
and the synthesized beam was about 30\arcsec $\times$30\arcsec, 
slightly different for different sources.  
The primary beam was about 6\arcmin~at 33.75 GHz.
All correlators to the most distant antenna from the center of 
array were unstable during the observations in 1997.  
The corrupt baselines were carefully flagged out.
The extra track on B1 was obtained in order
to observe extended emission around the edge of the primary beam 
in the previous observation.
In this track, a northern field, B1(N), and a southwestern field, B1(SW),
were observed for 15 min each sequentially.
The distances from the phase centers of each field to the center of
the previous observation were less than half of the primary beam.
BIMA was in a special compact configuration for this observation.
Baseline range was about the same as D configuration, but because
the compact array had more shorter baselines,
it produced a 43\arcsec$\times$36\arcsec~beam for B1.

In order to resolve the extremely narrow linewidth of the quiescent
dark clouds, the correlators were set in the mode with
the highest resolution available on BIMA, 0.054 \km/s.
In this mode, we only had one spectral window in each sideband.
The total bandwidth of each spectral window was 6.25 MHz (55.3 \km/s).

The 1-cm receivers operated from 27 to 36 GHz with a single
sideband response at the bottom and top of the tuning range and with
double sideband response in the center.  
Because the lower sideband was 2--3 times more sensitive than the upper
sideband, we set the observed line in the lower sideband
and the upper sideband was out of the operating range. As a result,
we obtained a single sideband system.
The system temperatures ranged from 40K to 96K during the 
observations in 1997
and from 130K to 230K for the observation in 1998.  The much higher system 
temperature in the second year was due to worse weather.

We chose the brightest quasars within 10$^\circ$ of the sources as 
the phase calibrators. The phase calibrators were observed for 5 minutes
every 30 minutes on the sources.
We also observed a planet for 5 minutes for each source in order to calibrate 
flux.
Because the observed bandwidth was very narrow
and the signal to noise ratios were not high for the short observation time for
each source, passband calibration was unnecessary.

All data reduction was carried out using the MIRIAD software
(Sault, Teuben, \& Wright 1995).
Bad visibility data were flagged, and data were phase calibrated 
and flux calibrated before being transformed into images.
For those sources with line detections, the continuum was subtracted
in the $u-v$ plane, although the continuum was within the noise level.
Natural weighting was used for constructing the images, and
the CLEAN algorithm was used to deconvolve the dirty beam from the image.
The boundary of the regions for deconvolution were chosen approximately 
between -1 $\sigma$ and 1 $\sigma$ rms levels around the sources in dirty maps, 
which provided better images especially for sources with 
high-sidelobe dirty beams.  The rms noise level of all sources in Table 1 were
estimated from a region in the clean map which did not contain signal.
Primary beam correction was applied to the three fields of B1 which were
then combined linearly.
Taper was applied on the combined map of B1 to avoid excessive noise 
amplification at the edge of the mosaiced field and to achieve approximately
uniform noise across the image.

\section{Results and Analysis}
Out of the 11 objects, 7 are detected with strong
CCS emission and extremely narrow linewidthes (0.1--0.8 \km/s).
Among the 4 objects without CCS detection (DR21, OMC-N4, S106, W3),
all except OMC-N4 are active star-forming regions, which suggests that
CCS could be destroyed in the later stage of star formation.
OMC-N4 is a dense ($\sim 10^6$ cm$^{-3}$) region 
at the position north of the OMC1 core without active star formation
where N bearing molecules are especially abundant.
The 7 clouds with CCS detection show tremendous clumpy structure. 
We identified the clumps in these CCS clouds and analyzed them 
separately.
A clump is identified in channel maps if there is an isolated peak with
intensity greater than 3 $\sigma$ in the image-velocity data cube 
across more than 3 channels. 
We treat B335 and L63 as one clump each.
B335 has double peaks, but it was previously modeled as 
a collapsing cloud (Zhou et al.\ 1993).
L63 has the lowest signal to noise ratio, although it may
separate into two clumps at the 2 $\sigma$ level.

The peaks of individual clumps are listed in Table 2, and
channel maps (Fig.\ 1--7(a)) show the clumps marked 
with a letter designation.
The $V_{lsr}$ and linewidth of each clump 
are fitted with a Gaussian at the peak
position and are listed in Table 2.  
68\% of clumps have total linewidths $\leq$ 0.25 \km/s.
If we exclude B1 and B335, both containing point-like IRAS sources,
only one clump out of 16 has a total linewidth $>$ 0.25 \km/s.
Assuming the kinetic temperature T$_k$ = 10 K,
7 clumps with nonthermal linewidths $<$ 0.1 \km/s are among
the most quiescent clumps that have been found.
The integrated intensity maps are shown in Fig.\ 1--7(b).
Fig.\ 1(c) shows the central velocity map of B1, and Fig.\ 2--7(c) show 
interesting position-velocity diagrams for the rest of clouds.
The distribution of CCS is discussed in \S 4.4.
Fig.\ 8 shows the line profile of each clump.

The brightness temperature of the intensity peak of each clump in Table 2
is obtained from the Rayleigh-Jeans Law,

\begin{equation}
T_b = 13.6\times \frac{I(Jy/beam)~\lambda_{mm}^2}{\theta_a~\theta_b}~(K),
\end{equation}

\noindent where $\theta_a$ and $\theta_b$ are the dimensions of the synthesized 
beams listed in Table 1.   
To calculate the optical depth and the total column density, we need the
excitation temperature  $T_{ex}$ of CCS.
The multiple-line CCS survey of SYO
determined $T_{ex}$ = 5\,K for B1, B335, L63, and L1544, and $T_{ex}$ = 5.2\,K
for L1498.  We adopt $T_{ex}$ = 5\,K for all our samples except L1498. 
Because of the high intensity of L1498, its excitation temperature must be 
greater than 5.5\,K unless the optical depth is negative.   
We adopt a value $T_{ex}$ = 6\,K from Benson et al.\ (1998). 
Therefore, the optical depth $\tau$ in Column 5 of Table 2 
was obtained from the radiative transfer equation

\begin{equation}
\tau = -ln(1-J(T_b)/J(T_{ex})),
\end{equation}

\noindent where $J(T)$ is the Plank function.
The total column density of CCS in Column 6 of Table 3 was calculated from

\begin{equation}
N(CCS) = 2.37\times 10^{-9}~\frac{Q(T_{ex})~exp(E_l/kT_{ex})~W}{(2J+1)~\theta_a~\theta_b~A}~(cm^{-2}),
\end{equation}

\noindent where $Q$($T_{ex}$) is the partition function, 
$E_l$ is the energy of lower level, 
$A$ is the Einstein A-coefficient for the transition,
and $W=\int I_\nu d\nu$ is the integrated line intensity. 
We adopted frequencies of transitions and the Einstein A-coefficient 
from WLG : $Q$ (5\,K) = 23.8017,  $Q$ (6\,K) = 30.5570,
$E_l$ = 1.6061\,K, and $A$ = 1.5999\,$\times$\,10$^{-6}$\,s$^{-1}$.

The virial masses in Column 7 of Table 3 were determined by assuming
a self-gravitating sphere with $n\propto r^{-1}$ and a gaussian velocity 
distribution for each clump.
MacLaren et al. (1988) derived the virial masses of such clouds as

\begin{equation}
M_{vir}=190~R~\Delta V_{H_2}^2~(M_\odot),
\end{equation}

\noindent 
where $R$ is the cloud radius in pc and $\Delta V_{H_2}$ is 
the line-of-sight FWHM speed of H$_2$ molecules in \km/s.
If the density law is between $n\propto r^{-2}$ and $n\propto constant$,
the virial mass would only change from 0.6 to 1.1 $M_{vir}$
as given in equation (4). 
The slope of the \N(CCS)--$n_{H_2,vir}$ correlation derived in
\S 4.2 is independent of the assumed density law. 
The radii of the clumps R listed in Columns 4 and 5 of Table 3 
are estimated from half of the geometric average
of the FWHM of major and minor axes of the clumps.  
Assuming the nonthermal part of the CCS and H$_2$ linewidth 
are the same, $\Delta V_{H_2}$ can be calculated from 
$\Delta V_{CCS}$ and the difference between their thermal widths.

\begin{eqnarray}
\Delta V_{H_2}^2 
&=& 8ln(2)\frac{kT_k}{m_{H_2}}+\Delta V_{NT}^2 \nonumber \\
&=& \Delta V_{CCS}^2 + 8ln(2)kT(\frac{1}{m_{H_2}}-\frac{1}{m_{CCS}}),
\end{eqnarray}

\noindent 
where $T_k$ is the kinetic temperature. We adopt $T_k$=10\,K 
for all clouds.  Because the observed $\Delta V_{CCS}$ is 
only a few channels wide for some clumps,
the correction for the convolution effect is necessary.  
The true $\Delta V_{CCS}$ is estimated from 
$(\Delta V_{CCS,true})^2 = (\Delta V_{CCS,obs})^2-0.054^2$. 
If we assume all the clumps are in viral equilibrium, 
we can calculate the H$_2$ density from the virial mass 
and estimate the fractional abundance of CCS, 
X(CCS) = $N_{CCS}/(2R\cdot n_{H_2,vir})$.  
The results are listed in Column 8 and 9 of Table 3 and 
range from 0.1--9$\times$10$^{-10}$, which brackets 
the value $4.9\times 10^{-10}$ obtained from the LVG calculation
for L1498 by WLG.
Large uncertainties for $M_{vir}$, H$_2$ density, 
and the fractional abundance come from the unknown dimension 
along the line of sight.  If the estimated dimension is 
larger by a factor of 2, then the $M_{vir}$ would be 
larger by the same factor, the H$_2$ density would be 
smaller by a factor of 4, and the fractional abundance of 
CCS would be larger by a factor of 2.

\section{Discussion}

\subsection{Individual Clouds}

\subsubsection{B1}
B1 is an extensive molecular cloud in the Perseus complex. 
A distance 350 pc is adopted for comparison with other results,
although there is some possibility that B1 may be located 
in the other molecular component
at less than 200 pc along the line of sight toward Perseus
(\cite{cbd85}). 
Physical properties of the envelope and main core of B1 
have been determined by the multiple-line study of 
Bachiller et al.\ (1990).  In their study,
the main core traced by \NH3 has a linewidth of 0.8 \km/s 
and a density greater than $10^4$ cm$^{-3}$. 
There are two IRAS sources (IRAS 03301+3057 and IRAS 03304+3100)
in the observed fields (Fig.\ 1). 
These two source may play a role in 
destroying the physical condition for CCS to exist.

We identify eight clumps in the channel map (Fig.\ 1).
Clump A--F correspond to the NH$_3$ main core, and
the CCS distribution of the main core is very similar to 
the NH$_3$ distribution observed by Bachiller et al.\ (1990).
The total mass of the main core estimated from the sum of 
the virial masses of clump A--F is $13.6 M_{\odot}$, 
which is much less than that derived from NH$_3 (\sim
150 M_{\odot}$) (\cite{bdm90}).  This can be understood
if \NH3 traces the densest region of the core and 
CCS traces the envelope of the core (see \S4.4).
A systematic velocity field with a range from 6.0 \km/s to 7.1 \km/s
is found along an arc formed by clumps A, B and F.  This appears to be
a rotational motion about IRAS 03301+3057.  
The arc also has larger linewidths than that of other clumps in B1,
and the linewidths of all clumps are smaller than the width of \NH3 lines.
B1(N) and B1(SW) both have almost coherent central velocity.
The large velocity differences between them and 
the nearest ends of the main core indicate that
these two clumps are kinematically distinct from the main core system.

\subsubsection{B133}
B133 is a Bok globule at a distance of less than 400 pc;
it has an IRAS source, IRAS 19035-0657,
at $\sim$2.5\arcmin~east of the center.
Four clumps can be identified in the peak intensity map (Fig.\ 2).
Clump A is weaker and has a narrower linewidth than the other clumps.
Clumps B, C, and D may spatially connect to each other, and clump A
may be isolated from them. 
Clump B has double peaks, which may be an infall feature or 
just simply two velocity components along the line of sight.
The position-velocity diagram along C--B2--D shows a slight velocity gradient,
about 1 \msarcsec = 2.5$\times 10^{-3}$ \msAU.
The FWHM also increases from C, B, and to the southern east edge of
clump D.

\subsubsection{B335}
B335 is a low-mass star-forming Bok globule at a distance of 250 pc.  
A cold, compact far-infrared source lies in its center 
(\cite{kdh83}), and it also shows dynamical features of a collapsing cloud
with a bipolar outflow (\cite{fl82}) and infall (\cite{zek93}).
The observed asymmetric line profiles have been modeled as inside-out collapse
(\cite{zek93}, \cite{ceg95}). Velusamy et al.\ (1995) combined VLA and
DSN 70\,m single-dish observations of 
the 22 GHz CCS $J_N=2_1-1_0$ transition and show that the
distribution of CCS emission is non-spherical and clumpy with depletion
in the central 10\arcsec.  
They conclude that CCS is characteristic of early astrochemical time
(t$\le 10^5$ yr) and more prominent in the region
just outside the infall envelope.

The source shape in our integrated intensity map of B335 is 
similar to that of the DSN 70\,m map of Velusamy et al.
The central position of the collapsing region (\cite{zek93})
is offset toward the east of
the integrated intensity peak by half of our synthesized beam.
We obtain an average X(CCS)$\sim 1.6\times 10^{-10}$, which
is consistent with the values from the LVG calculation by Velusamy et al.\
: X(CCS)$\sim 3\times 10^{-9}$ in the outer envelope and
X(CCS)$\sim 5\times 10^{-11}$ in the central depletion region.
The dimension of the region with blue asymmetric line profiles 
can be estimated from the position-velocity
diagrams along the major and minor axes;  
it is $\sim$ 100\arcsec$\times30\arcsec$, 
corresponding to 0.12 pc\,$\times$\,0.04 pc,
which is larger than the CCS depletion region.

\subsubsection{L63}
L63 is an isolated globule at a distance of 160 pc.
A G0 type star is located $\sim$2.5\arcmin~southwest of the center.
If this star is associated with the cloud, it may explain why
L63 is the weakest CCS source in our sample.
Two condensations at different velocity can be separated 
in the position-velocity diagram.  
The western one is centered at 5.8 \km/s, 
and the eastern one at 5.9 \km/s. 
There is a velocity gradient along the east-west direction,
$\sim$ 1.4~\msarcsec~=~8.7 $\times 10^{-3}$~\msAU.
The linewidth of the peak we obtained, 0.167 \km/s, 
is narrower than that obtained by
SYO (0.8 \km/s) and \cite{bcm98} (0.36 \km/s).

\subsubsection{L183}
L183, or L134N, is a part of the L134-L183-L1778 system of 
interstellar clouds at high galactic altitude, b$\sim 36^{\circ}$.  
No far-infrared source has been detected toward L183 (\cite{svn83}).
Clark \& Johnson (1981) observed OH 6 cm and H$_2$CO 2 cm lines 
for this system, and suggested the retrograde rotation of 
these cores can be explained by magnetic braking.

Three clumps are identified. Each clump is in a different velocity range.
The flux of clump A could be stronger than the value in Table 2,
because the intensity peak of clump A lies on the edge of the primary beam.
Clump B has a different shape in channel maps, and it may divide
into more clumps if the signal to noise ratio of the map were higher.
The velocity gradient along A, B is obvious in the position-velocity diagram,
with value 1.2 \msarcsec = 7.5 $\times 10^{-3}$ \msAU.

\subsubsection{L1498}
L1498 is an extremely quiescent pre-protostellar core in the Taurus complex.
It has been studied extensively with
multiple transitions of CCS at 22 GHz, 45 GHz, and 94 GHz observed 
by WLG.
Our L1498 map clearly shows two concentrations in 
the position-velocity diagram along the major axis,
which is consistent with the map obtained by WLG.
The peanut-like distribution in the p-v diagram shows 
the two concentrations are at about the same velocity of 7.87 \km/s
and have a similar linewidth of 0.18 \km/s.
The virial mass of each clump is $\sim$ 1.4 $M_{\odot}$, and the 
H$_2$ density derived from the virial mass is $\sim$ 3$\times 10^5$ cm$^{-3}$.
The fractional abundances of CCS are 1.2 and 1.5$\times 10^{-10}$ for 
the NW and SE clumps, and both values are 
less than that obtained from the LVG calculation
by WLG , 4.9$\times 10^{-10}$. 

\subsubsection{L1544}
L1544 is a starless core in the Taurus complex at 140 pc.
It is not associated with any point-like source.
IRAS 05013+2505, peaked at 100\,$\mu$m, is nearby, 
but is extended over 5\arcmin.
Tafalla et al.\ (1998) did a multiline study of L1544
and found that a large area in L1544 has a double peak spectrum.
They explained this feature
as an infall with high self-absorption in the center.
They modeled the velocity structure with a simple two-layer 
collapsing model and concluded that the data are inconsistent 
with the inside-out collapse model.  
They also ruled out ambipolar diffusion driven collapse 
because the speed of ionized 
species is similar to that of neutrals.

Six clumps are identified in L1544,
which is elongated along the northwest to southeast direction.
The position-velocity digram along the major axis of this elongated cloud
shows a ring structure with an inner hole of 0.04 pc\,$\times$\,0.13 \km/s 
and an outer boundary of 0.16 pc\,$\times$\,0.55 \km/s.
This ring structure can be interpreted 
as a collapsing disk or ring almost edge-on with infall velocity
$\gtrsim$0.1 \km/s.  The rotation velocity can be estimated from
the velocity difference between the upper and bottom edge
of the ring, which is only $\sim$ 0.05 \km/s, comparable to 
our spectral resolution.  
Because the self-absorption toward the center has been observed with CS,
the central hole of the ring should not indicate a physical hole 
in the center. Instead it may be a further evidence that
CCS only traces the outer envelope of dense cores.

\subsection{Correlation of \N(CCS) with Density}

It has been widely suggested that CCS is more abundant
in the early stage of protostellar cores. However, observationally
it is difficult to determine a clear inverse correlation 
of \N(CCS) with H$_2$ density as an indicator of the cloud age.
Except for the work done by Fuente et al.\ (1990), 
observational data show that the correlation is fairly weak 
(SYO, WLG, Benson et al.\ 1998). 
Fuente et al.\ (1990) show a positive correlation 
between \N(CCS) and \N(\C18O) for seven dark clouds,
which seems inconsistent with the prediction of chemical
models unless the seven clouds are all in a very early stage.

Since CCS distributions are clumpy and may be shell-like around
the higher density core whereas CO apparently exists in a larger
area, the depth of the CCS column should be a few factors smaller
than that of the CO column.  Therefore, the average densities obtained
from CO may not be able to represent the density of CCS clumps.
In addition, based on the complex distributions and 
active chemical reactions ( formation in the outer lower 
density region and destruction in the central higher 
density region), it seems unsuitable to use a single 
average density to represent the chemical behavior of CCS
in the whole cloud.  Therefore, it is necessary to 
analysis the \N(CCS)--$n$(H$_2$) correlation at the clump level.

In order to have a systematic analysis at the clump level,
we adopt the virial density to represent the H$_2$ densities.
If the clouds are not virialized, the H$_2$ densities will be
incorrect, but the statistical conclusion would still be valid
if the virial densities were scaled from the actual densities
with similar factors.  We obtain a very weak correlation, 
$N$(CCS) $\sim 10^{13.7\pm 1.0}~n_{H_2}^{-0.25\pm 0.18}$ cm$^{-2}$ 
(Fig.\ 9).  This relation brackets the weak correlation,
$N$(CCS) $\sim 10^{14.1}~n_{H_2}^{-0.22}$ cm$^{-2}$ obtained by
WLG for one of the three velocity components of TMC-1D.  
In their study, the correlations for the other 
two components of TMC-1D are too weak to be distinguished 
from a flat distribution.  Therefore, we conclude that 
even though CCS is more abundant in the early stage
of a contracting cloud, the abundance of CCS is not likely to
trace the age of the contracting cores very sensitively.
Moreover, an almost flat distribution could imply that the time for
creating and destroying CCS may be very short compared to
the lifetime of CCS.

\subsection{Cloud Equilibrium}
We compare the average H$_2$ column density obtained
from C$^{18}$O or NH$_3$ observations (Myers et al.\ 1983,
Benson \& Myers 1989) to
the average virial H$_2$ column density of each cloud
for our CCS data.
The average virial column density of each cloud is calculated by
averaging that of all clumps in the cloud.
The comparison results are listed in Table 4.
Except for cloud B1, the observed column density is much smaller than
the CCS virial column density, which suggests that the CCS clumps
are not virialized. Since the real densities of the CCS clumps should be
smaller than the virial densities,
X(CCS) estimated in Table 3 can only provide lower limits.

\subsection{Distribution of CCS}
The channel maps of our sampled clouds (Fig.\ 1--7(a)) show the
remarkable clumpy structure of the CCS distribution.
As suggested by Velusamy et al.\ (1995) from the observation of B335,
the clumpy structure may originate from the destruction of 
CCS in the densest region of the pre-protostellar cores.
Therefore, combined with our conclusion in \S 4.3, 
we suggest that the clumps traced by CCS may not be self-contracting
cores for protostars, but only higher density condensations 
in the envelope of a collapsing core.

The above idea can be further supported by
comparing the relative positions of CCS and higher density regions, 
traced by \NH3 emission or submillimeter continuum sources.  
We marked \NH3 peaks with X symbols and marked 
submillimeter continuum sources with ellipses in 
the integrated intensity maps (Fig.\ 1--7(b)).
The comparison results are listed in Table 4.
We found that except for cloud B1, 
CCS generally exists around the higher density regions.
This can be seen obviously for B133, L63, and L183, where
the CCS emission exists asymmetrically around 
the submillimeter continuum sources.
For L1498, the \NH3 peak appears between two CCS clumps, and
for L1544, the \NH3 peak might be in the center of the ring structure
discussed in \S 4.1.7.
The \NH3 peak in B335 is close to the integrated intensity peak,
and close to the place with central depletion reported by Velusamy et al.\ (1995).
Although for the B1 main core (clump A--F) the CCS distribution is similar
to the \NH3 distribution observed by Bachiller et al.\ (1990),
the CCS distribution appears to be more clumpy than the \NH3 distribution.
Also, the \NH3 peak is at the gap between clump B and E.
Therefore, the CCS distributions for all clouds are consistent with
the idea that CCS traces the envelope of pre-protostellar cores
rather than being individual self-gravitation cores themselves.

\subsection{Linewidth-Radius Relations}

Almost 2 decades after Larson (1981) discovered a clear correlation between
the linewidths and the cloud sizes $\Delta v \propto R^{0.38}$,
the physical origin of this Larson's Law remains an important unsolved
problem.   It has been widely discussed that
turbulence or magnetohydrodynamic turbulence may provide 
the large nonthermal component of linewidths and the scaling relation through 
energy dissipation, though a complete theory has not yet been established.

A question has been asked: does the relation terminate at small scale?
Goodman et al.\ (1998) proposed a scenario that a transition to coherent
linewidths takes place at the length scale $\sim$ 0.1 pc
based on their observations of \NH3 lines with spatial resolution 60\arcsec
-80\arcsec.
The nonthermal components of linewidths in coherent cores are
less than the thermal widths of H$_2$, but strictly non-zero.
They suggest that the coherent length scale may come from the dissipation
threshold of MHD waves.

The CCS clouds in our sample with physical sizes comparable to 
the coherent length scale can be a test ground of the existence of 
the coherent cores.
The velocity dispersion of each pixel can be easily calculated 
from the second-order moment analysis.
A cut-off level of 2 $\sigma$ on the spectrum must be set 
when calculating the velocity dispersion
in order to reduce the confusion from noise.
However, this high cut-off level can significantly reduce
the velocity dispersion if the signal-to-noise ratio is low.
Assuming an ideal Gaussian profile, we calculated how the correction coefficient
varies with the signal-to-noise ratio.
Averaging the corrected linewidth maps, 
we found that the linewidth-radius
relation for each cloud is very consistent with a flat distribution.
The result is not sensitive to the center position we chose.
An example of L1498 is shown in Fig.\ 10.
Since the linewidths are almost homogeneous everywhere for each cloud,
the widths from the Gaussian fitting listed in Table 2 
can represent the typical linewidth in the cloud.  
The total linewidth of CCS is almost 
contributed by the nonthermal width, since CCS is a heavy molecule.
The thermal linewidth of H$_2$ at T$_k=$10\,K is about 0.5 \km/s.
Therefore, the nonthermal linewidths of our sampled clouds are
comparable or less than the thermal linewidth of H$_2$.

The observed facts seem to support the picture of coherent cores.  
However, as we discussed in \S4.4, we have to be cautious that
CCS may only trace the envelope rather than the center of cores.
Therefore, the linewidth at small radius on the sky does not necessarily 
represent the linewidth at small distance from the center of the core.
Nevertheless, we can say at least the linewidths are very homogeneous
in the narrow density regions traced by CCS.

\section{Summary}
CCS is an abundant molecule in quiescent dark clouds
and is not detectable in active star-forming regions.
Seven out of 25 clumps have extremely small nonthermal linewidths, 
$<$ 0.1 \km/s.
The distributions of CCS are very clumpy, and the CCS clumps
are not necessarily symmetric around the higher density regions
where protostars may be born.
The CCS clumps themselves are not likely to collapse into protostars,
but rather exist transiently in the regions with appropriate 
densities as the cloud evolves.  Therefore, we conclude that CCS traces
pre-protostellar cores in the very early stage of star formation
or traces the envelope of pre-protostellar cores in the later stage.
Homogeneous linewidths in CCS clouds may provide evidence that
the nonthermal linewidths transition to coherence 
at the length scale $\sim$ 0.1 pc,
which is proposed by Goodman et al.\ (1998).
Because CCS is a Zeeman molecule and also an excellent kinematic probe,
follow-up polarimetry and high velocity resolution studies could provide
a good opportunity to understand
the physical properties of star-forming regions in the early stage.
L1544, with a very systematic velocity field and strong CCS emission,
is an excellent target to model the physical condition of a collapsing core.

\acknowledgments
Research at the Laboratory for Astronomical Imaging
was supported by NSF grant AST 96-13999 and by the University of Illinois.


%
%
%
%

\clearpage



%




\clearpage

%
%

%

\clearpage

\figcaption{(a) CCS channel maps of B1 obtained from 3 mosaic fields. The rms noise level varies across the maps and ranges from
0.1 \Jb~at the center of the central field (R.A.(J2000)=03:33:17.83, Dec.(J2000)=31:07:29.80)
to the maximum value of 0.19 \Jb~beyond 3\arcmin~from the center.
Contours are drawn at -0.5, -0.3, 0.3, 0.5, 0.8, 1.2 \Jb.
The black square and star indicate the positions of IRAS 03301+3057 
and 03304+3100 (=LkH$\alpha$327).
(b) The integrated intensity map of B1.  Contours are drawn at 
-0.35, -0.25, -0.15, 0.15, 0.25, ..., 0.75 \Jbk.  
The X symbol marks the \NH3 peak (Bachiller et al.\ 1990).
(c) The intensity weighted velocity of B1. Contours are at 6.2, 6.5, 6.8, 7.1 
\km/s.  }

\figcaption{(a) CCS channel maps of B133. The 1-$\sigma$ rms noise level
is 0.33 \Jb, and contours are drawn at -3, 3, 5, 7, 9 $\sigma$ levels.
(b) The integrated intensity map of B133.  The 1-$\sigma$ rms noise level is 0.1
2 \Jbk, and contours are drawn at -3, 3, 6 $\sigma$ levels. The X symbol marks the \NH3 peak (Benson \& Myers 1989), and the circle marks the submillimeter continuum source (Ward-Thompson et al.\ 1994).
(c) The position-velocity diagram along the cut shown in (b).  Contours are at
15\%-90\% of the peak, and the contour interval is 15\%.}

\figcaption{(a) CCS channel maps of B335. The 1-$\sigma$ rms noise level
is 0.22 \Jb, and contours are drawn at -3, 3, 5, 7, 9 $\sigma$ levels.  The dot
marks
the position of the spectrum shown in Fig 8.
(b) The integrated intensity map of B335. The 1-$\sigma$ rms noise level is 0.10
 \Jbk, and contours are drawn at -3, 3, 5, 7 $\sigma$ levels. The X symbol mark
 the \NH3 peak (Benson \& Myers 1989), and the plus symbol marks the center of CCS depletion (Velusamy et al.\ 1995).
(c) The position-velocity diagram along the cut shown in (b). Contours are at
30\%-90\% of the peak, and the contour interval is 15\%.}

\figcaption{
(a) CCS channel maps of L63. The 1-$\sigma$ rms noise level is 0.22 \Jb, and contours are drawn at -2, 2, 3, 5 $\sigma$ levels.  The dot marks the position of the spectrum shown in Fig 8.
(b) The integrated intensity map of L63. The 1-$\sigma$ rms noise level is 0.035 \Jbk, and the contour is drawn at 3 $\sigma$ level. The X symbol marks the \NH3 peak (Benson \& Myers 1989), and the ellipse marks the submillimeter continuum source (Ward-Thompson et al .\ 1994).
(c) The position-velocity diagram along the cut shown in (b). Contours are at 30\%-90\% of the peak, and the contour interval is 15\%.}

\figcaption{
CCS channel maps of L183. The 1-$\sigma$ rms noise level is 0.37 \Jb, and contours are drawn at -3, 3, 5 $\sigma$ levels.  The plus symbol marks the pointing center, and clump A is on the boundary of the primary beam.
(b) The integrated intensity map of L183. The 1-$\sigma$ rms noise level is 0.035 \Jbk, and contours are drawn at -6, -3, 3, 6, 9, 12 $\sigma$ levels. The X symbol marks the \NH3 peak (Benson \& Myers 1989), and the ellipse marks the submillimeter continuum source (Ward-Thompson et al.\ 1994).
(c) The position-velocity diagram along the cut shown in (b). Contours are at 30\%-90\% of the peak, and the contour interval is 15\%.}

\figcaption{
(a) CCS channel maps of L1498. The 1-$\sigma$ rms noise level is 0.15 \Jb, and contours are drawn at -5, -3, 3, 5, 8, 11, 14, 17, 20, 23, 26 $\sigma$ levels.  The dots mark the positions of the spectrum shown in Fig 8.
(b) The integrated intensity map of L1498. The 1-$\sigma$ rms noise level is 0.050 \Jbk, and contours are drawn at -6, -3, 3, 6, 9, 12, 15, 18 $\sigma$ levels. The X symbol marks the \NH3 peak (Benson \& Myers 1989).
(c) The position-velocity diagram along the cut shown in (b).  Base contour is at 3-$\sigma$ rms noise level, 0.45 \Jb, and the contour interval is also 0.45 \Jb.}

\figcaption{
(a) CCS channel maps of L1544. The 1-$\sigma$ rms noise level is 0.13 \Jb, and contours are drawn at -6, -3, 3, 6, 9, 12, 15, 18 $\sigma$ levels.
(b) The integrated intensity map of  L1544. The 1-$\sigma$ rms noise level is 0.050 \Jbk, and contours are drawn at -3, 3, 6, 9, 12 $\sigma$ levels. The X symbol marks the \NH3 peak (Benson \& Myers 1989).
(c) The position-velocity diagram along the cut shown in (b).  Base contour is at 2-$\sigma$ rms noise level, 0.26 \Jb, and the contour interval is also 0.26 \Jb. Note that the central two contours are at 0.26 and 0.52 \Jb, which indicate a hole in the center.}

\figcaption{ Line profiles of CCS clumps.}

\figcaption{
CCS column density vs. virial H$_2$ density for all clumps (see \S 4.2).  Note that B1D is an unresolved clump and L183A is on the edge of the primary beam.}

\figcaption{
The linewidth-radius relation of L1498. The central position is chosen at the intensity peak.  The rise feature at R$>$0.1 pc can be washed out by choosing different central positions, however, the flatness within R=0.1 pc remains unchanged.}

%

\clearpage

{\scriptsize
\begin{deluxetable}{lcrcccccc}
\tablecaption{Observation parameters}
\tablehead{ 
\colhead{}&	\colhead{}& \colhead{}& \colhead{Observing}& \colhead{rms}&\multicolumn{3}{c}{synthesized beam} \\
\cline{6-8}
\colhead{Source}  & \colhead{R.A.(J2000)}   & \colhead{Dec.(J2000)}& 
\colhead{time (hr)} &\colhead{(Jy/beam)}& \colhead{major(\arcsec)}&\colhead{minor(\arcsec)} & \colhead{P.A.($^{\circ}$)} & \colhead{Detection} 
}
\startdata
B1      & 03:33:17.83   &31:07:29.80  & 5.0 & 0.12 & 32 & 25 & -72& Y\\
B1(N)   & 03:33:17.83   &31:09:30.00  & 3.0 & 0.19 &43 & 36 & ~10 & Y\\
B1(SW)  & 03:33:11.00   &31:05:50.00  & 3.0 & 0.19 &43 & 37 & ~10 & Y\\
B133    & 19:06:08.90   &-06:52:20.00 & 3.0 & 0.33 & 39 & 27 & -1& Y\\
B335    & 19:37:00.99   &07:34:16.80  & 5.0 & 0.22 & 32 & 24 & ~70& Y\\
DR21    & 20:39:01.30   &42:19:40.80  & 6.7 & 0.21 & 38 & 30 & ~71 & N \\
L63     & 16:50:13.90   &-18:06:22.00 & 7.6 & 0.15 & 39 & 32 & -13  & Y\\
L183    & 15:54:09.20   &-02:51:07.00 & 3.0 & 0.37 & 42 & 24 & -75& Y\\
L1498   & 04:10:51.50   &25:10:16.00  & 5.3 & 0.15 & 32 & 31 & ~85& Y\\
L1544   & 05:04:17.20   &25:10:44.00  & 4.7 & 0.13 & 32 & 29 & ~24& Y\\
OMC-N4  & 05:35:16.80   &-05:19:31.00 & 4.0 & 0.44 & 56& 22 & -72 & N  \\
S106-CN & 20:27:29.60   &37:22:54.00  & 2.6 &0.23 &36 & 29  & ~68 & N \\
W3      & 02:25:36.70   &62:06:11.00  & 3.7 &0.19 &39 & 22 &  -78 & N \\
\enddata
\end{deluxetable}

\clearpage
 
\begin{deluxetable}{lccclrr}
\tablecaption{Line parameters}
\tablehead{
\colhead{Source} & \colhead{Distance\tablenotemark{a}} &
\colhead{$I_{peak}$} & \colhead{$T_b$} &
\colhead{$\tau_{peak}$\tablenotemark{b}} & \colhead{$V_{lsr}$}
& \colhead{$\Delta v$} \\
\colhead{} & \colhead{(pc)}& \colhead{(Jy/beam)} & \colhead{(K)} &
\colhead{} & \colhead{(km/s)} & \colhead{(km/s)}}
 
\startdata

B1A  & 350 & 1.36 & 1.61 & 0.48 & 6.11$\pm$0.01 & 0.42$\pm$0.03 \\
B1B  &     & 0.73 & 0.87 & 0.23 & 6.78$\pm$0.03 & 0.78$\pm$0.08 \\
B1C  &     & 0.81 & 0.96 & 0.26& 6.48$\pm$0.02 & 0.24$\pm$0.04 \\
B1D  &     & 0.70 & 0.83 & 0.22& 6.50$\pm$0.01 & 0.13$\pm$0.03 \\
B1E  &     & 1.02 & 1.21 & 0.34& 6.89$\pm$0.01 & 0.29$\pm$0.03 \\
B1F  &     & 0.72 & 0.85 & 0.23& 7.22$\pm$0.03 & 0.61$\pm$0.06 \\
B1(N) &     & 1.23 & 1.46 & 0.42& 6.97$\pm$0.01 & 0.38$\pm$0.03 \\
B1(SW) &     & 2.12 & 2.51 & 0.90& 6.81$\pm$0.01 & 0.34$\pm$0.02 \\
B133A  &400& 1.87  & 1.90    & 0.61&11.96$\pm$0.01 & 0.13$\pm$0.03 \\
B133B1 &   & 3.03  & 3.14 & 1.4~ & 12.10$\pm$0.01 & 0.15$\pm$0.03 \\
B133B2 &   & 3.16  & 3.28 & 1.5~ & 12.27$\pm$0.01 & 0.09$\pm$0.02 \\
B133C  &   & 2.43  & 2.52 & 0.91 & 12.22$\pm$0.01 & 0.20$\pm$0.03 \\
B133D  &   & 2.54  & 2.63 & 0.97 & 12.37$\pm$0.02 & 0.46$\pm$0.04 \\
B335   &250& 1.96  & 2.69    & 1.13&8.45$\pm$0.01 & 0.41$\pm$0.03 \\
L63    &160& 0.91 & 0.79   & 0.21  &5.90$\pm$0.01 & 0.16$\pm$0.02\\
L183A&150& 2.05 & 1.79 & 0.55 & 2.40$\pm$0.01 & 0.11$\pm$0.02 \\
L183B&& 2.23 & 1.95 & 0.62 & 2.49$\pm$0.01 & 0.15$\pm$0.02 \\
L183C&& 1.50 & 1.31 & 0.37 & 2.71$\pm$0.02 & 0.15$\pm$0.04   \\
L1498A &140& 4.32 & 4.77 & 2.4~ & 7.87$\pm$0.01 & 0.17$\pm$0.01 \\
L1498B &   & 3.93 & 4.34 & 1.8~ & 7.86$\pm$0.01 & 0.17$\pm$0.01 \\
L1544A &140& 2.39 & 2.79 & 1.1~ & 7.09$\pm$0.01 & 0.18$\pm$0.01 \\
L1544B    && 2.03 & 2.37 & 0.82 & 7.12$\pm$0.01 & 0.12$\pm$0.01\\
L1544C    && 1.67 & 1.95 & 0.62& 7.19$\pm$0.01&  0.11$\pm$0.01\\
L1544D    && 1.77 & 2.06 & 0.67& 7.33$\pm$0.01& 0.25$\pm$0.01\\
L1544E    && 2.44 & 2.85 & 1.1~& 7.35$\pm$0.01& 0.20$\pm$0.01 \\
L1544F    && 2.06 & 2.41 & 0.84&7.41$\pm$0.01& 0.12$\pm$0.01 \\
\tablenotetext {a}{Reference: Hilton \& Lahulla (1995).}
\tablenotetext {b}{$T_{ex}$ = 6K for L1498; $T_{ex}$ = 5K for the rest of clouds.}
\enddata
\end{deluxetable}
\clearpage

\begin{table*}
\begin{center} {\small Table 3. Derived quantities}        \\
\vskip 0.2cm
\begin{tabular}{lcccccccc}
\hline\hline
(1) & (2) &(3) &(4) &(5) &(6) &(7) &(8) &(9)  \\
Source & Major & Minor & R & R & $N(CCS)$       & $M_{vir}$    & $n_{H_2,vir}$ & $ X(CCS) $\\
       & Axis(\arcsec) & Axis(\arcsec) & (\arcsec) & (pc)&$(10^{12}cm^{-2})$& $(M_{\odot})$& $(10^4cm^{-3})$& $(10^{-10})$\\
\hline
B1A &~45 & 40 & 20 & 0.068 & 5.3 & 2.2 & 27 & 0.93 \\
B1B & ~75 & 55 & 30 & 0.11 & 4.9 & 3.6 &10& 1.4 \\
B1C & 140 & 25& 30 & 0.10~ & 1.3 & 3.3 &12&0.33\\
B1D & ~25 & 15 & 10 & 0.034 & 0.33 & 1.1 & 108 & 0.029 \\
B1E & ~40 & 40 & 20 & 0.068 & 5.0 & 2.2 &27&0.88\\
B1F & ~50 & 40 & 20 & 0.076 & 2.6 & 2.5 & 21&0.51\\
B1(N)&110& 55 & 40 & 0.14~ & 2.9 & 4.4 & 6.8 & 1.0 \\
B1(SW) & ~80&70& 40 & 0.13~ & 6.6 & 4.1 & 7.7 & 2.2 \\
B133A & ~50 &50& 25 & 0.097 & 0.52 & 2.2&9.3&0.19\\
B133B1 & ~90&45& 30 & 0.13~ & 3.4 & 2.9&5.6&1.6\\
B133B2 & ~80&50& 30 & 0.13~ & 2.2 & 2.7 & 5.3 & 1.0 \\
B133C & ~50 &40& 20 & 0.087 & 3.4 & 2.1&13&1.0\\
B133D & ~90&50 & 35 & 0.13~ & 6.6 & 5.2&10&1.7\\
B335 & 120&50 & 40 & 0.097 & 7.3 & 3.6 &15&1.6\\
L63 & 150&70 & 50 & 0.08~ & 0.84 & 1.8&15&0.24\\
L183A & ~90&40 & 30 & 0.044 & 0.21 & 1.0 & 45 &0.11\\
L183B &110&50 & 40 & 0.055& 1.6 & 1.3&30.2&0.32\\
L183C & ~50&30 & 20 & 0.029 & 1.1 & 0.7 & 106& 0.11 \\
L1498A& 100&70& 40 & 0.057 & 7.1 & 1.4 & 27.6 & 1.5 \\
L1498B& 100&60& 40 & 0.054 & 5.4 & 1.3 & 31.9 & 1.0 \\
L1544A& ~65 &55& 30 & 0.041 & 3.0 & 1.0 & 56 & 0.42 \\
L1544B& ~50 &50& 25 & 0.033 & 2.2 & 0.8 & 80 & 0.26 \\
L1544C& ~45 &35& 20 & 0.027 & 0.96 & 0.6 & 125 & 0.096 \\
L1544D& ~55 &40& 25 & 0.031 & 3.2 & 0.7 & 99 & 0.34 \\
L1544E& ~50 &35& 20 & 0.027 & 4.4 & 0.6 & 125 & 0.41 \\
L1544F& ~45 &45& 20 & 0.031 & 2.5 & 0.7 & 99 & 1.1 \\

\hline
\end{tabular}
\end{center}
\end{table*}
}
\clearpage

\begin{table*}
\begin{center}
Table 4. Comparison between CCS clouds. \\
\begin{tabular}{lcccc}
\hline\hline
&(1)&(2)&(3)&(4)\\
Source &Correspondence & Double peaks & \N(H$_2$),$_{NH_3 or C^{18}O}$ & \N(H$_2$),$_{virial}$ \\
       & NH$_3$, continuum&& (10$^{21}$cm$^{-2}$) & (10$^{21}$cm$^{-2}$)\\
\hline
B1(Main)  & Yes & No& 24 & 36 \\
B133 & No & Yes& 0.86& 29\\
B335 & No & Yes& 2.8& 44\\
L63  & No & No&8.3 & 36\\
L183 & No & No&9.5 & 53\\
L1498& No & No& 5.7& 49\\
L1544& No & Yes& 13& 59\\
\hline
\end{tabular}
\end{center}
\end{table*}


\begin{thebibliography}{}
\bibitem[Bachiller et al.\ 1990]{bdm90} Bachiller, R., Del Rio Alvarez, S., \& Menten, K. M. 1990, \aap, 236, 461
\bibitem[Benson \& Myers 1989]{bm89} Benson, P.J., \& Myers, P.C. 1989, ApJ. Suppl., 71, 89
\bibitem[Benson et al.\ 1998]{bcm98} Benson, P.J., Caselli, P., \& Myers, P.C. 1998, ApJ, 506, 743
\bibitem[Carlstrom et al.\ 1996]{cml96} Carlstrom, J.E., Marshall, J., Laura, G., 1996, ApJ, 456, L75
\bibitem[Cernicharo et al.\ 1985]{cbd85} Cernicharo,J., Bachiller, R., \& Duvert, G. 1985, \aa, 149,273
\bibitem[Choi et al.\ 1995]{ceg95} Choi, M., Evans, N.J., II, Gregersen, E.M., \& Wang, Y. 1995, ApJ, 448,742
\bibitem[Clark \& Johnson 1981]{cj81} Clark, F.O., \& Johnson, D.R. 1981, ApJ, 247, 104
\bibitem[Codella \& Scappini 1998] {} Codella, C. \& Scappini, F. 1998, \mnras, 298, 1092
\bibitem[Frerking \& Langer 1982]{fl82} Frerking, M.A., \& Langer, W.D. 1982, ApJ, 256, 523
\bibitem[Fuente Cernicharo Barcia \& Gomez-Gonzalez 1990]{}Fuente, A., Cernicharo, J., Barcia, A. \& Gomez-Gonzalez, J. 1990, \aap, 231, 151 
\bibitem[Goodman et al.\ 1998]{} Goodman, A. A., Barranco, J. A., Wilner, D. J. \& Heyer, M. H. 1998, \apj, 504, 223
\bibitem[Hilton \& Lahulla 1995]{}Hilton, J. \& Lahulla, J. F. 1995, \aaps, 113, 325 
\bibitem[Keene et al.\ 1983]{kdh83} Keene, J., Davison, J.A., Harper, D.A., Hildebrand, R.H., Jaffe, D.T., Lowenstein, R.F., Low, F.J., \& Pernic, R. 1983, ApJ, 274, L43
\bibitem[Kuiper et al.\ 1996]{klv96} Kuiper, T.B.H., Langer, W.D., \& Velusamy, T. 1996, ApJ, 468, 761
\bibitem[Larson 1981]{} Larson, R. B. 1981, \mnras, 194, 809
\bibitem[Myers Linke \& Benson 1983]{}Myers, P. C., Linke, R. A. \& Benson, P. J. 1983, \apj, 264, 517
\bibitem[Saito et al. 1987]{} Saito, S. , Kawaguchi, K., Yamamoto, S., Ohishi, M.  \& Suzuki, H.  1987, \apjl, 317, L115 
\bibitem[Sault et al.\ 1995]{stw95} Sault, R. J., Teuben, P. J., \& Wright, M. C. H. 1995, in ASP Conf. Proc. 77, Astronomical Data Analysis Software and Systems IV, ed. R. A. Shaw, H. E. Payne, \& J. J. E. Hayes (San Francisco: ASP), 433 
\bibitem[Sargent et al.\ 1983]{svn83} Sargent, A.I., van Duinen, R.J., Nordh, H.L., Fridlund, C.V.M., Aalders, J.W.G., Beintema, D. 1983, AJ, 88, 88
\bibitem[Scappini \& Codella 1996] {} Scappini, F. \& Codella, C. 1996, \mnras, 282, 587
\bibitem[]{} Shinnaga, H. \& Yamamoto, S. 1999, in preparation
\bibitem[Suzuki et al. 1984]{}Suzuki, H., Kaifu, N., Miyaji, T., Morimoto, M., Ohishi, M. \& Saito, S. 1984, \apj, 282, 197
\bibitem[Suzuki et al.\ 1992]{syo92} Suzuki, H., Yamamoto, S., Ohishi, M., Kaifu, N., Ishikawa, S.-I., Hirahara, Y. \& Takano, S. 1992, \apj, 392, 551 
\bibitem[Tafalla et al. 1998]{tmm98}Tafalla, M., Mardones, D., Myers, P. C., Caselli, P., Bachiller, R., \& Benson, P. J. 1998, \apj, 504, 900 
\bibitem[Velusamy et al.\ 1995]{vkl95} Velusamy, T., Kuiper, T.B.H., \& Langer, W.D. 1995, ApJ. Lett., 451, L75
\bibitem[Ward-Thompson et al.\ 1994]{} Ward-Thompson, D., Scott, P. F., Hills, R. E. \& Andre, P. 1994, \mnras, 268, 276 
\bibitem[Welch et al.\ 1996]{w96} Welch, W. J., et al. 1996, PASP, 108, 93 
\bibitem[Wolkovitch et al.\ 1997]{wlg97} Wolkovitch, D., Langer, W.D., Goldsmith, P.F., \& Heyer, M. 1997, ApJ, 477, 241
\bibitem[Zhou et al.\ 1993]{zek93} Zhou, S., Evans, N.J., II, K$\ddot{o}$mpe, C., \& Walmsley, C.M. 1993, ApJ, 404, 232 
\end{thebibliography}
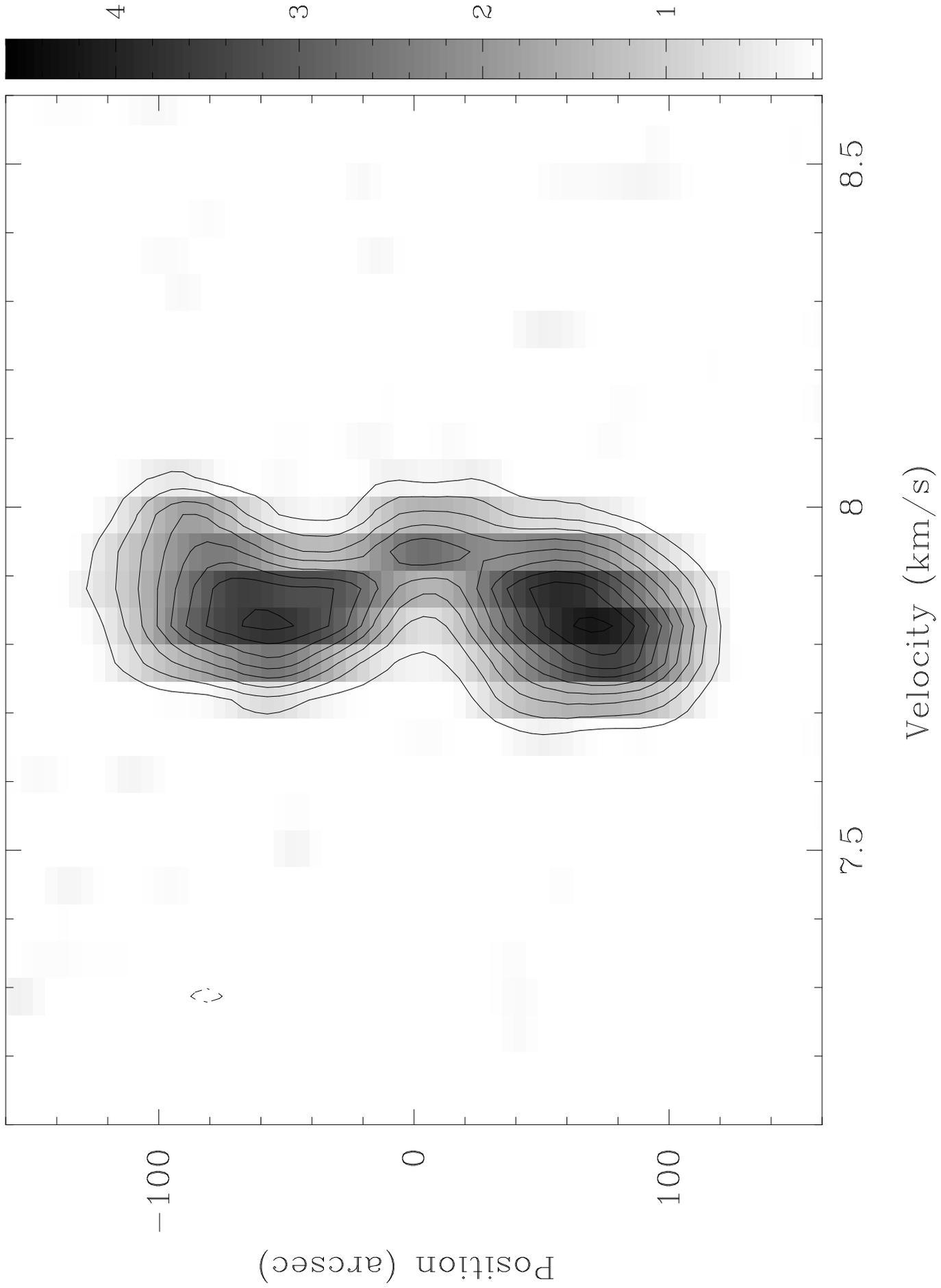
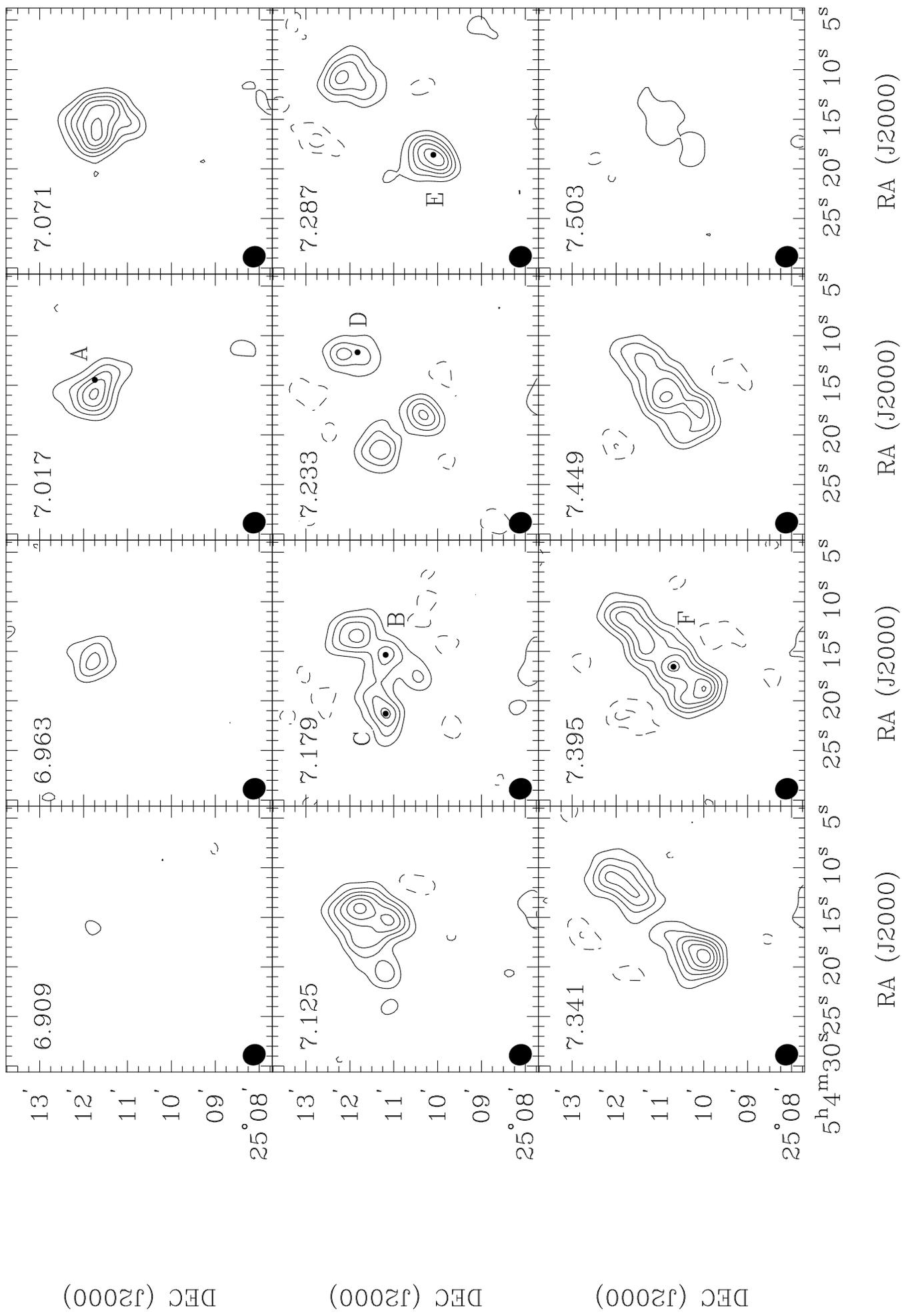
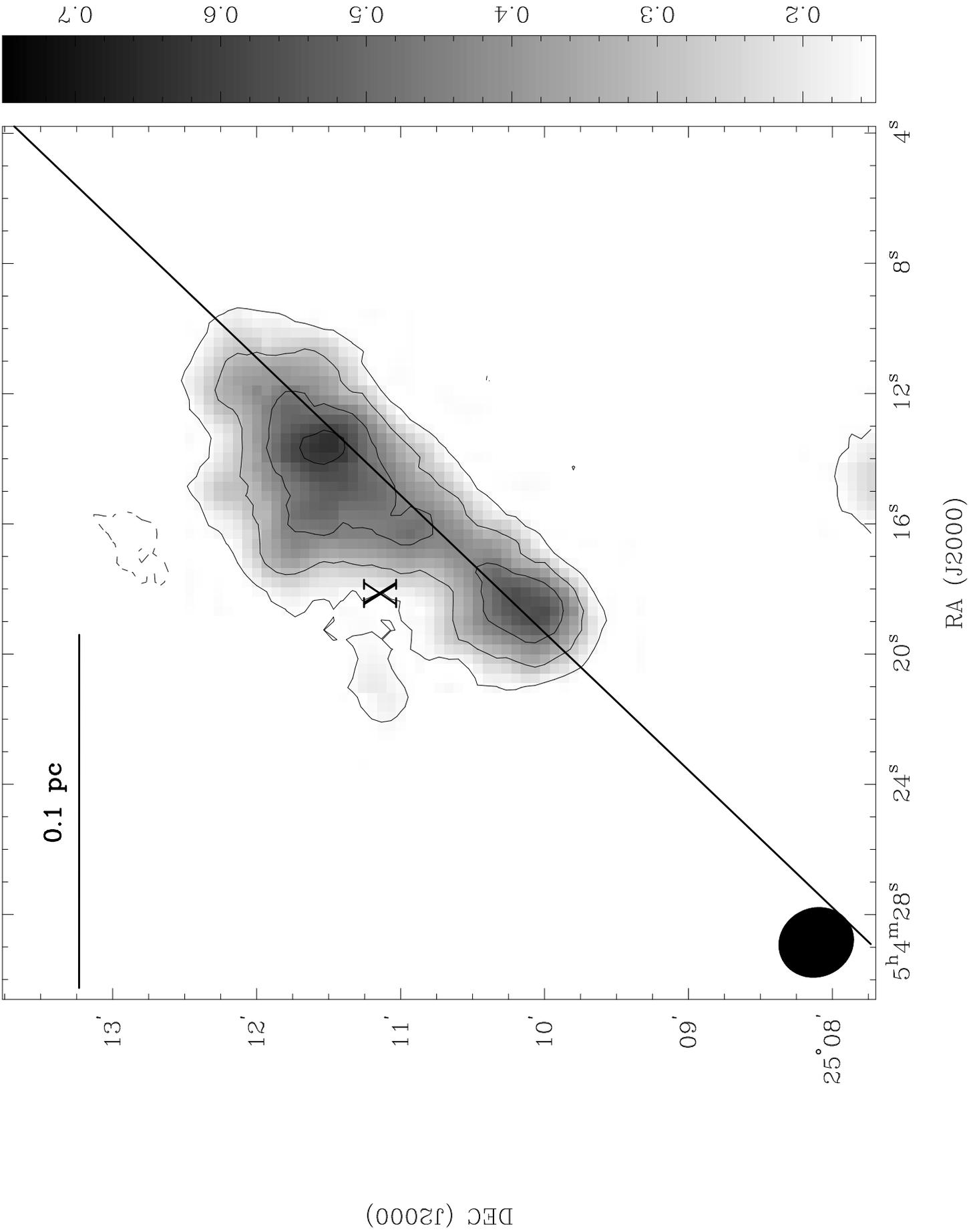
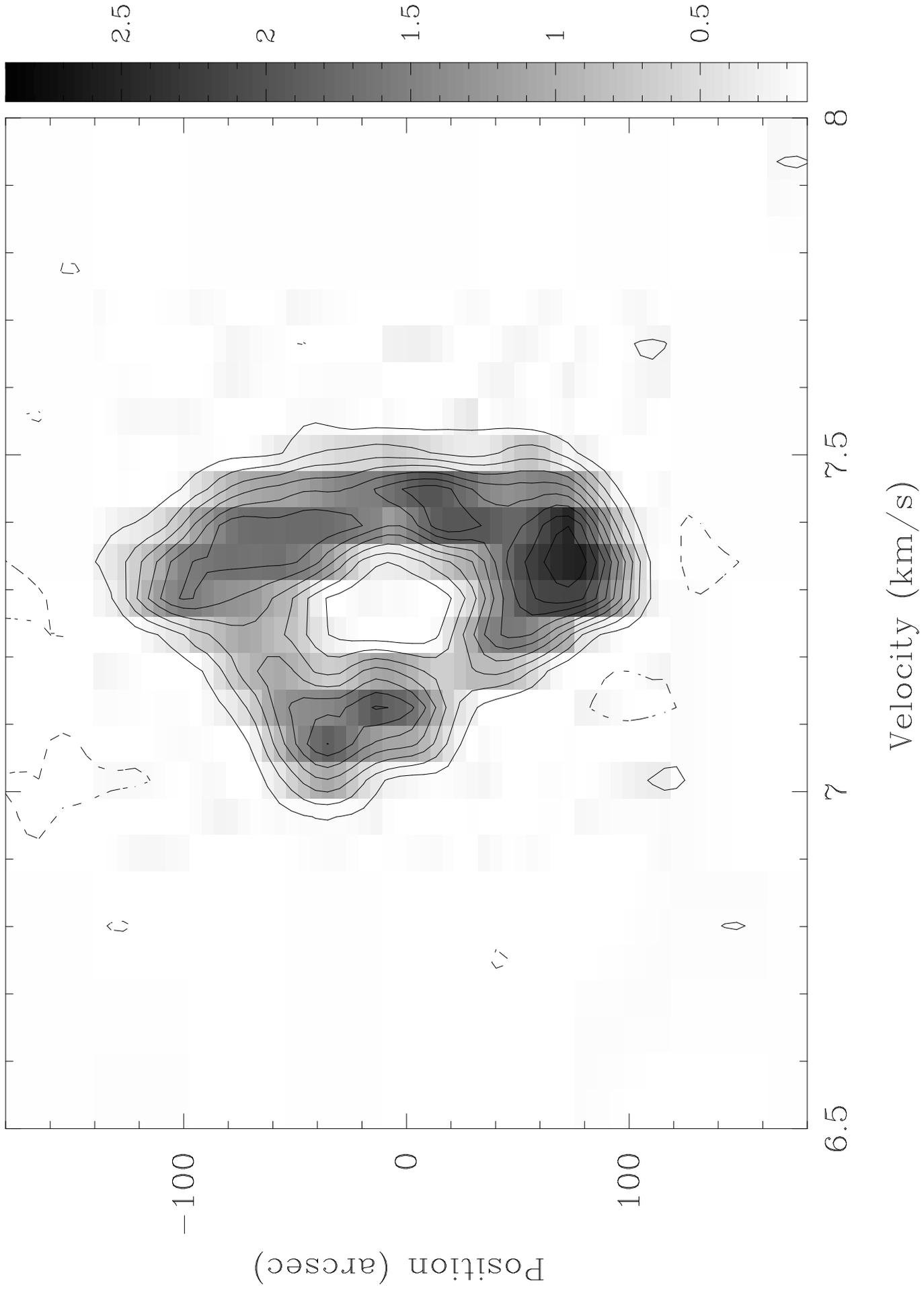
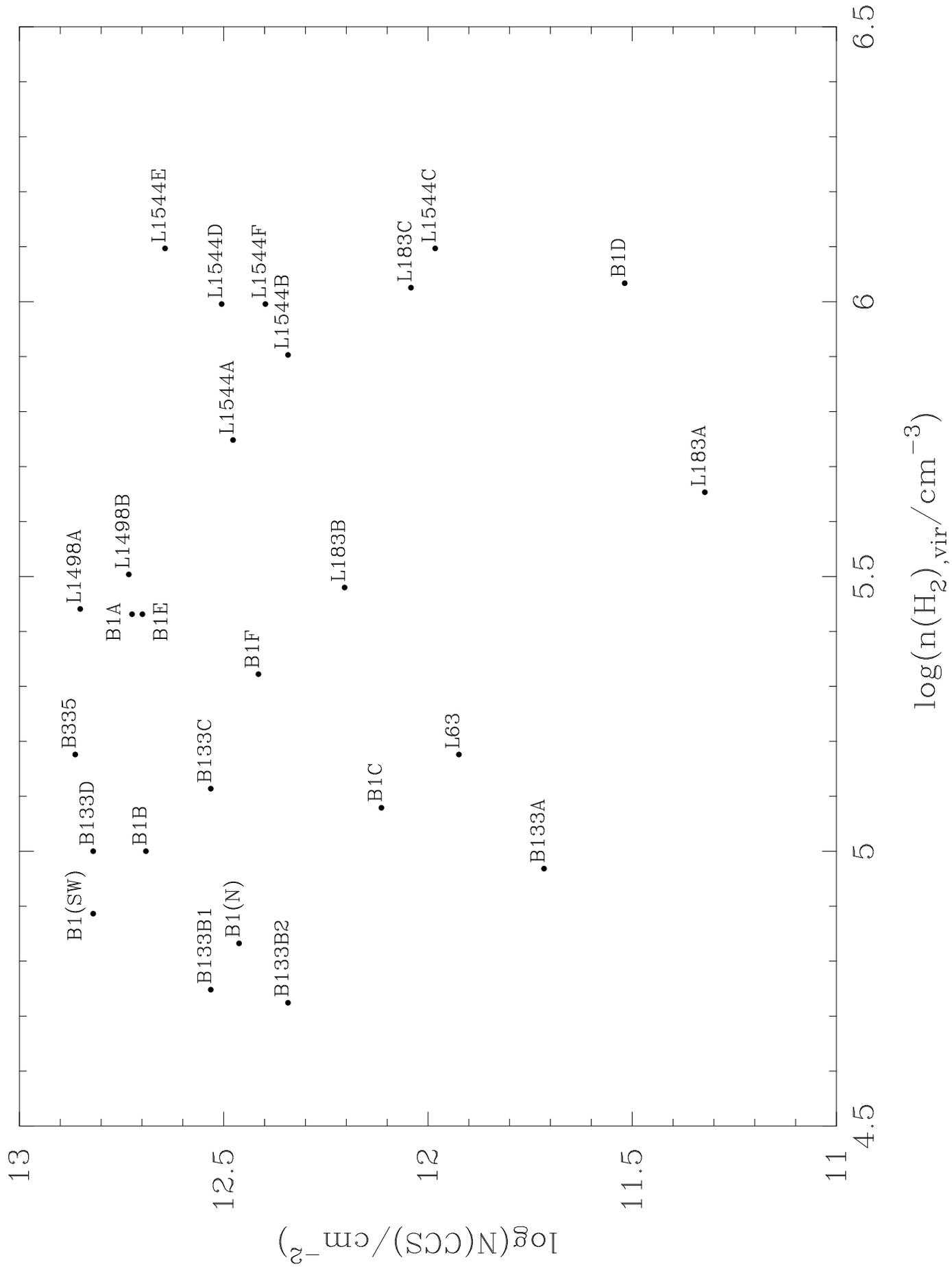
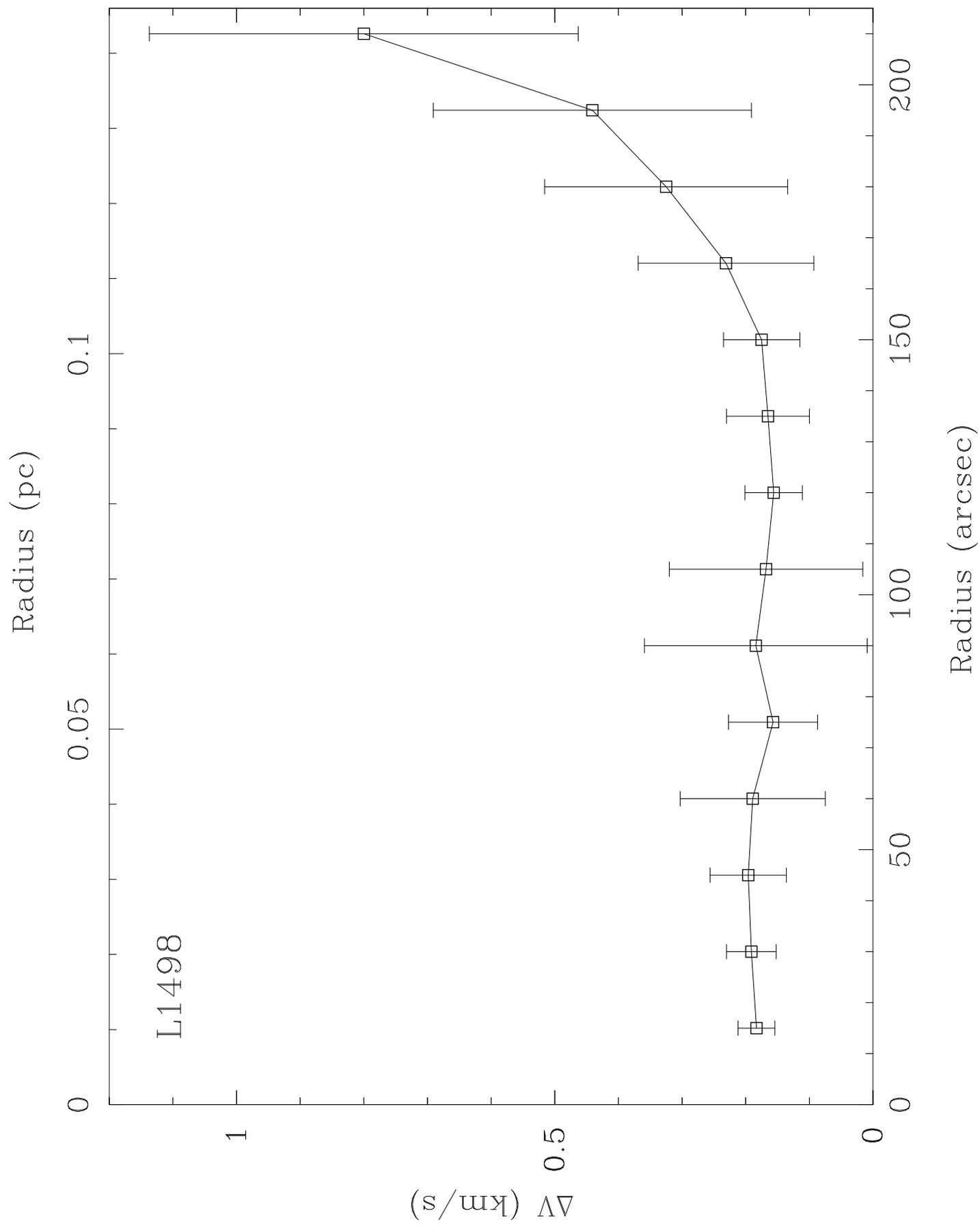
\end{document}